\documentclass[10pt,final,doublecolumn]{IEEEtran}
\usepackage{amsmath}
\usepackage{latexsym}
\usepackage{graphicx}
\usepackage{bbding}
\usepackage{indentfirst}
\usepackage{algorithm,algorithmic}
\usepackage{setspace}
\usepackage{float}
\usepackage{epstopdf}
\usepackage{amssymb}
\usepackage{amsfonts}
\usepackage{enumerate}
\usepackage{multicol}
\usepackage{color}
\usepackage{diagbox}
\usepackage{bm}
\usepackage{amssymb}
\usepackage{stfloats}
\usepackage{epstopdf}
\usepackage{threeparttable}
\usepackage{epstopdf}
\usepackage{threeparttable}
\usepackage{subfigure}
\usepackage{makecell}

\usepackage{cite}
\usepackage[table]{xcolor}

\usepackage[linewidth=0.5pt]{mdframed}
\IEEEoverridecommandlockouts
\allowdisplaybreaks[4]

\begin{document}
\title{Reconfigurable Coupler Antenna for Wireless Networks
}
	\author{{Xiaodan Shao, \IEEEmembership{Member, IEEE}, Chuangye Shan, Weihua Zhuang, \IEEEmembership{Fellow, IEEE}, Xuemin (Sherman) Shen, \IEEEmembership{Fellow, IEEE}}
	\thanks{X. Shao, W. Zhuang, and X. Shen are with University of Waterloo, Waterloo, ON N2L 3G1, Canada.}
}
\maketitle

\IEEEpeerreviewmaketitle

\begin{abstract}
The reconfigurable coupler antenna (RCA), also called the flexible coupler antenna (FCA), is a new technique that aims to improve the performance of wireless communication networks by reconfiguring the positions and rotations of low-cost couplers around fixed-position active antennas to harness mutual coupling.
Specifically, different couplers can independently adjust their positions and/or rotations at the transceiver to reshape the induced currents on the couplers for radiation, thereby collaboratively achieving mechanical beamforming for directional signal enhancement or nulling. The position and/or rotation reconfiguration of passive couplers provides a new and cost-effective means of enhancing wireless communication performance, while significantly reducing the antenna and radio-frequency (RF) chain costs of conventional active arrays. The compact and low form-factor structure of the RCA makes it particularly appealing for devices with stringent size, weight, and power (SWAP) constraints. In this article, we provide an overview of RCA to reveal its promising capabilities in wireless networks, including its system modeling, practical implementation, and competitive advantages over existing techniques. We present a variety of RCA-enabled  performance enhancements in terms of mechanical beamforming gain, path-loss reduction, fading mitigation, spatial multiplexing gain, interference suppression, and geometric gain. Furthermore, we elaborate on the design challenges of RCA as well as promising solutions, and discuss the key applications of RCA in wireless networks. Finally, numerical results are presented to verify the substantial capacity gains enabled by RCA-aided transmission in wireless networks.
\end{abstract}

\section{Introduction}
In future sixth-generation (6G) networks, the demand for massive connectivity, ultra-low latency, and high energy efficiency drives the exploration of advanced wireless communication techniques \cite{10054381}. Multiple-input multiple-output (MIMO) technology plays an essential role in 6G. MIMO performance gains scale with the number of antennas. However, since traditional MIMO requires each antenna to be fed by a unique radio-frequency (RF) chain, deploying more antennas requires larger installation spaces.
Such a requirement increases the hardware cost of the transceiver and makes MIMO systems costly and bulky, which remain crucial unresolved issues. For example, migrating massive MIMO from sub-6 GHz to millimeter-wave (mmWave) bands typically demands more sophisticated processing and higher-cost hardware. These issues motivate the pursuit of flexible and cost-efficient solutions for future wireless networks.

Recently, the reconfigurable coupler antenna (RCA)), also referred to as the flexible coupler antenna (FCA), has been proposed as a promising new technique for reconfiguring the wireless propagation environment by controlling coupler positions and/or rotations \cite{shao2026coupler,ROCATWC,RCATWC}. By dispensing with the requirement that each antenna be connected to a dedicated RF chain, an RCA consists of a single RF-fed active antenna and multiple couplers. The passive couplers can be independently adjusted in terms of their positions and/or rotations, while the position and rotation of the active antenna remain fixed. Here, the coupler denotes a passive element that interacts with the active antenna through near-field electromagnetic coupling and can function analogously to a reflector in Yagi-Uda antennas \cite{6605533}, rather than a conventional multiport RF circuit coupler designed to couple power between guided transmission lines \cite{1132816}.

By exploiting coupler movement and harnessing mutual coupling between the active antenna and nearby couplers, the couplers in the RCA radiate through the excitation induced by the active antenna, thereby enabling a new form of position-dependent mechanical beamforming.
The transceiver can be designed to control the coupler positions according to the spatial channel distribution to maximize system capacity. 
In contrast to existing wireless link adaptation techniques at the transmitter or receiver, RCA reshapes the wireless channel by adjusting highly controllable coupler positions. This provides a new degree of freedom (DoF) for performance enhancement and enables a reconfigurable wireless channel. 
By eliminating transmit RF chains for passive couplers and adjusting them only within a small local region on the order of the wavelength, RCAs offer a favorable tradeoff between geometric compactness and performance gain. Consequently, RCAs feature a compact and lightweight structure with low power consumption, which makes them suitable for small-sized terminals and practical deployment.

It is worth noting that the RCA system considered in this article is fundamentally different from existing six-dimensional movable antenna (6DMA) systems \cite{shao20246d,6dma_dis} and conventional electronically steerable parasitic array radiators (ESPARs) \cite{10993454}. Compared with 6DMA, in which active antennas connected to RF chains are translated and/or rotated \cite{10883029}, RCA moves passive couplers within a small wavelength-scale region while keeping the active antenna and its RF chain fixed. The radiated signals of RCAs are generated through near-field electromagnetic (EM) coupling, requiring no additional RF chains. Therefore, compared with 6DMA, RCA reduces the number of RF chains, hardware cost, and physical size. Consequently, RCAs can be installed not only on base stations (BSs) but also on space-limited terminals. Moreover, unlike electronically tunable ESPARs, RCA reconfigures passive couplers through mechanical position/rotation adjustment rather than impedance tuning. This provides a wider tuning range and enables RCA to approach the performance of a fully active array, at the cost of higher hardware complexity and  slower response. In addition, RCA bandwidth depends on frequency-dependent electrical spacing, mutual impedance, impedance matching, and radiation-pattern stability. Table~I summarizes the comparison among the RCA, 6DMA, ESPAR, and Yagi-Uda antenna, where $B$ denotes the number of active antennas in 6DMA. For a fair comparison, $B$ is set equal to the sum of the active antenna and passive couplers in the RCA.
\newcommand{\tabincell}[2]{\begin{tabular}{@{}#1@{}}#2\end{tabular}}
\begin{table*}[!t]
	\small
	\caption{Comparison of reconfigurable coupler antenna with other related technologies.}
	\centering
	\begin{tabular}{
			>{\columncolor{blue!15}}c
			>{\columncolor{black!10}}c
			>{\columncolor{blue!15}}c
			>{\columncolor{black!10}}c
			>{\columncolor{blue!15}}c
			>{\columncolor{black!10}}c
			>{\columncolor{blue!15}}c
			>{\columncolor{black!15}}c
		}
		\bfseries \tabincell{c}{Architecture}
		& \bfseries \tabincell{c}{Tunable\\parameter}
		& \bfseries \tabincell{c}{No. of\\RF chains}
		& \bfseries \tabincell{c}{Physical\\ compactness}
		& \bfseries \tabincell{c}{Performance\\gain}
		& \bfseries \tabincell{c}{Hardware \\cost}
				& \bfseries \tabincell{c}{Operating 
					 \\mechanism}
					\\
		\Xhline{1pt}
		\tabincell{c}{Reconfigurable coupler\\ antenna \cite{shao2026coupler,ROCATWC}}
		& \tabincell{c}{Position/rotation\\of passive coupler}
				& 1
		& \tabincell{c}{Compact}
		& High
		& Low
		& \tabincell{c}{	Passive coupling,\\ channel adaptation}
		\\
		\Xhline{0.5pt}
		6DMA \cite{shao20246d,6dma_dis}
		& \tabincell{c}{Position/rotation\\of active antenna}
		& $B$
		& \tabincell{c}{Moderate}
		& Very high
		& Medium
		& \tabincell{c}{Active radiation,\\ channel adaptation}
		\\
		\Xhline{0.5pt}
		\tabincell{c}{ESPAR/Yagi-Uda\\ antenna \cite{10993454,10818494}}
		& \tabincell{c}{Reflector/director \\ or load}
		& 1
		& Very compact
		& Low
		&Low
		& \tabincell{c}{Pattern control}
		\\
		\Xhline{1pt}
	\end{tabular}
	\label{Table1}
\end{table*}


\section{Signal Model and Hardware Architecture}
In this section, we first provide the general signal model for the RCA, and then discuss its mechanical beamforming and hardware architecture.

\subsection{Signal Model}
As shown in Fig.~\ref{circuit}, the RCA under consideration has a fixed-position active antenna and multiple passive couplers. The passive couplers can move within a designated area around the active antenna, and their positions can be mechanically adjusted with the aid of drive components \cite{shao2026coupler,RCATWC}. The active antenna is fed by a single RF chain. Signals radiated by the couplers arise from induced currents excited through near-field EM coupling, thereby eliminating the need for additional RF feeds. With a small local movement region for passive couplers, RCA improves wireless communication performance without moving the active antenna/RF chain, thereby reducing mechanical control complexity and energy consumption. This compact architecture makes RCA attractive for SWAP-constrained wireless terminals and platforms.

Mathematically, by stacking the active antenna and all passive couplers, the line-of-sight (LoS) geometric channel vector for the link between an RCA and a single fixed-position antenna can be written as
$\mathbf h(\mathbf p)=\gamma\big[1, \mathbf{a}^T(\mathbf p)\big]^T$ \cite{shao2026coupler},
where $\gamma$ denotes the common large-scale attenuation across all ports, $\mathbf p$ is the position vector of the couplers, $\mathbf a(\mathbf p)$ represents the LoS steering vector of couplers, and the leading entry $1$ corresponds to the normalized steering factor of the active antenna. In contrast to the conventional fixed-position antenna channel, the RCA-enabled wireless channel is dependent on coupler positions $\mathbf p$. Therefore, by adjusting the coupler positions, the transmitter or receiver can reshape the effective channel, thereby improving channel conditions and enhancing wireless system performance for different purposes. For example, when serving a coverage-limited user, the coupler positions can be optimized to coherently reinforce the received signal and improve the effective array gain, while in interference-limited scenarios, the coupler position control can create destructive combining toward undesired directions and thus suppress multiuser interference.

\subsection{Mechanical Beamforming}
\begin{figure*}[t!]
	\centering
		\setlength{\abovecaptionskip}{0.cm}
	\includegraphics[width=6.66in]{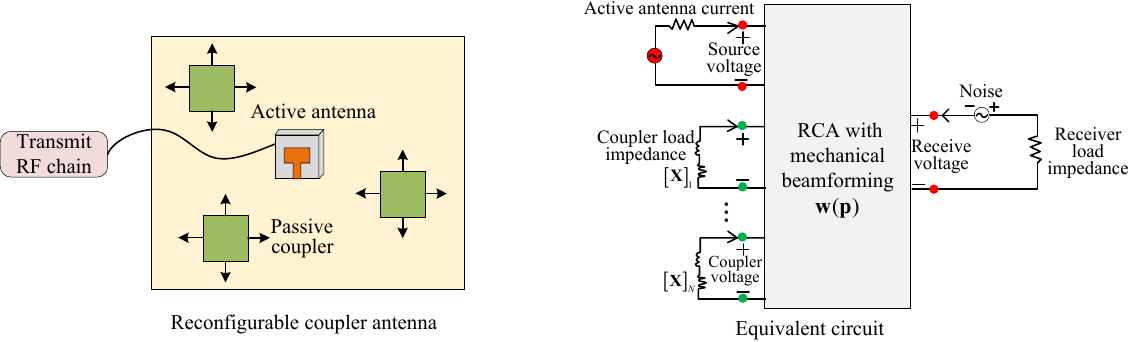}
	\caption{Architecture of RCA.}
	\label{circuit}
	\vspace{-0.59cm}
\end{figure*}

Mechanical beamforming in the RCA architecture shapes the radiation pattern by repositioning passive couplers surrounding the fixed active antenna. Specifically, adjusting the coupler position vector directly modifies the induced complex excitations across the antenna ports, which can be captured by position-dependent mechanical beamforming vector $\mathbf{w}(\mathbf{p})=(\mathbf{Z}(\mathbf{p})+\mathbf{X})^{-1}\bar{\mathbf{z}}(\mathbf{p})$, where $\mathbf{Z}(\mathbf{p})$ and $\bar{\mathbf{z}}(\mathbf{p})$ denote the mutual impedance matrix of all  couplers and the mutual impedance vector between the active antenna and all couplers, respectively, $\mathbf{X}$ denotes the load-impedance matrix of the couplers \cite{shao2026coupler,RCATWC}, as shown in the equivalent multiport circuit for the RCA in Fig.~\ref{circuit}.

In general, the effective channel between an RCA-antenna-based transmitter-receiver pair is determined by two main components, namely, the geometric propagation channel between the transmitter and receiver and the mechanical beamforming vector.
Mathematically, the received signal, $y$, is obtained by multiplying the incident signal, $s$, with the complex geometric channel and mechanical beamforming vector, i.e.,
$y= \mathbf h^T(\mathbf p){\mathbf w}(\mathbf p) s$.
Moreover, mechanical beamforming at each RCA can complement digital beamforming across the active antennas, thereby yielding a hybrid architecture that combines low-dimensional digital control with local coupler reconfiguration. 

\subsection{Hardware Architecture}
\begin{figure}[t!]
	\centering
	\setlength{\abovecaptionskip}{0.cm}
	\includegraphics[width=3.4in]{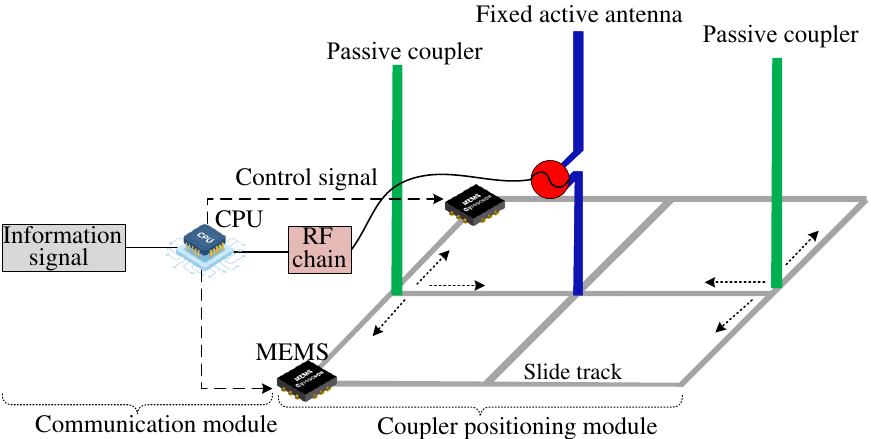}
	\caption{Hardware architecture for mounting the RCA at the transmitter/receiver.}
	\label{hardware}
	\vspace{-0.59cm}
\end{figure}

As shown in Fig.~\ref{hardware}, the RCA transceiver comprises a conventional communication module and a coupler positioning module. The communication module includes a single RF chain connected to a fixed active antenna and a central processing unit (CPU). The CPU performs baseband processing and generates control commands for coupler-position reconfiguration. For the coupler
positioning module, the coupler is installed on a three-dimensional (3D) mechanical slide, which is driven by micro-electromechanical systems (MEMS). With electrostatic actuation, MEMS devices can provide high positioning accuracy at low power consumption \cite{balanis2011modern}, which is well suited to wavelength-scale coupler motion. Active RF chains typically consume tens to hundreds of milliwatts, whereas MEMS-based actuation operates at the milliwatt level or consumes power only during actuation. Thus, the additional power for local MEMS actuation is much smaller than the RF-chain power saving enabled by the RCA architecture.

To reduce hardware complexity when the number of couplers is large, multiple couplers can be grouped into a subsurface module so that all elements in the same module translate or rotate jointly, thereby reducing the number of actuators and control lines. When moving couplers, a minimum distance between any pair of couplers, as well as between a coupler and an active antenna, is required to prevent physical overlap. Moreover, a minimum coupler-to-active spacing can avoid excessive near-field coupling and severe voltage standing wave ratio (VSWR) variation. Other ways to mitigate VSWR variation include impedance-matching-aware position optimization and a tunable matching network at the active antenna port.

\section{RCA Implementations}
 In this section, we discuss different ways to implement the RCA in wireless systems.

\subsection{RCA array}
Multiple RCAs can be deployed to form an RCA array, which enlarges the effective aperture and improves multiuser MIMO performance.
As shown in Fig.~\ref{implement}, an RCA array can adopt either a uniform linear array (ULA) or a uniform planar array (UPA) layout \cite{shao2026coupler}.
In a ULA RCA array, RCAs are arranged along one dimension to provide angular selectivity in the horizontal domain with low-complexity array-level beam steering.
In a UPA RCA array, RCAs are placed on a two-dimensional grid to enable joint azimuth-and-elevation beamforming and provide additional flexibility for elevation-domain interference management.

\subsection{General RCA array}
A general RCA array is illustrated in Fig.~\ref{implement}, where couplers are not pre-assigned to specific active antennas. Instead, all active antennas and passive couplers are deployed within a common region, and couplers can be repositioned around different active antennas.
The general structure enables flexible allocation of passive elements across active antennas. Compared with the ULA/UPA RCA array, it can reshape the effective aperture in a finer-grained manner to match heterogeneous spatial channel distributions, at the cost of increased optimization and control complexity.

\subsection{Dual-motion RCA}
RCAs typically allow coupler translations of several wavelengths while keeping the active antenna fixed. This limited range constrains path-loss mitigation and overall performance. Fig.~\ref{implement} illustrates a dual-motion RCA \cite{RCA}, which is an extended RCA architecture. 
In this extension, passive couplers translate on the wavelength scale to enable mechanical beamforming. Meanwhile, the active antenna slides along a rail over larger distances to reconfigure the transceiver geometry. When the active-antenna displacement is non-negligible relative to the signal propagation distance, the dual-motion RCA significantly improves link quality by reducing separation and facilitating LoS establishment. Mechanical beamforming provides extended angular coverage, which reduces the rails required for spatial coverage and lowers hardware costs.

\subsection{Rotatable Coupler Antenna}
\begin{figure*}[t!]
	\centering
		\setlength{\abovecaptionskip}{0.cm}
	\fbox{\includegraphics[width=6.4in]{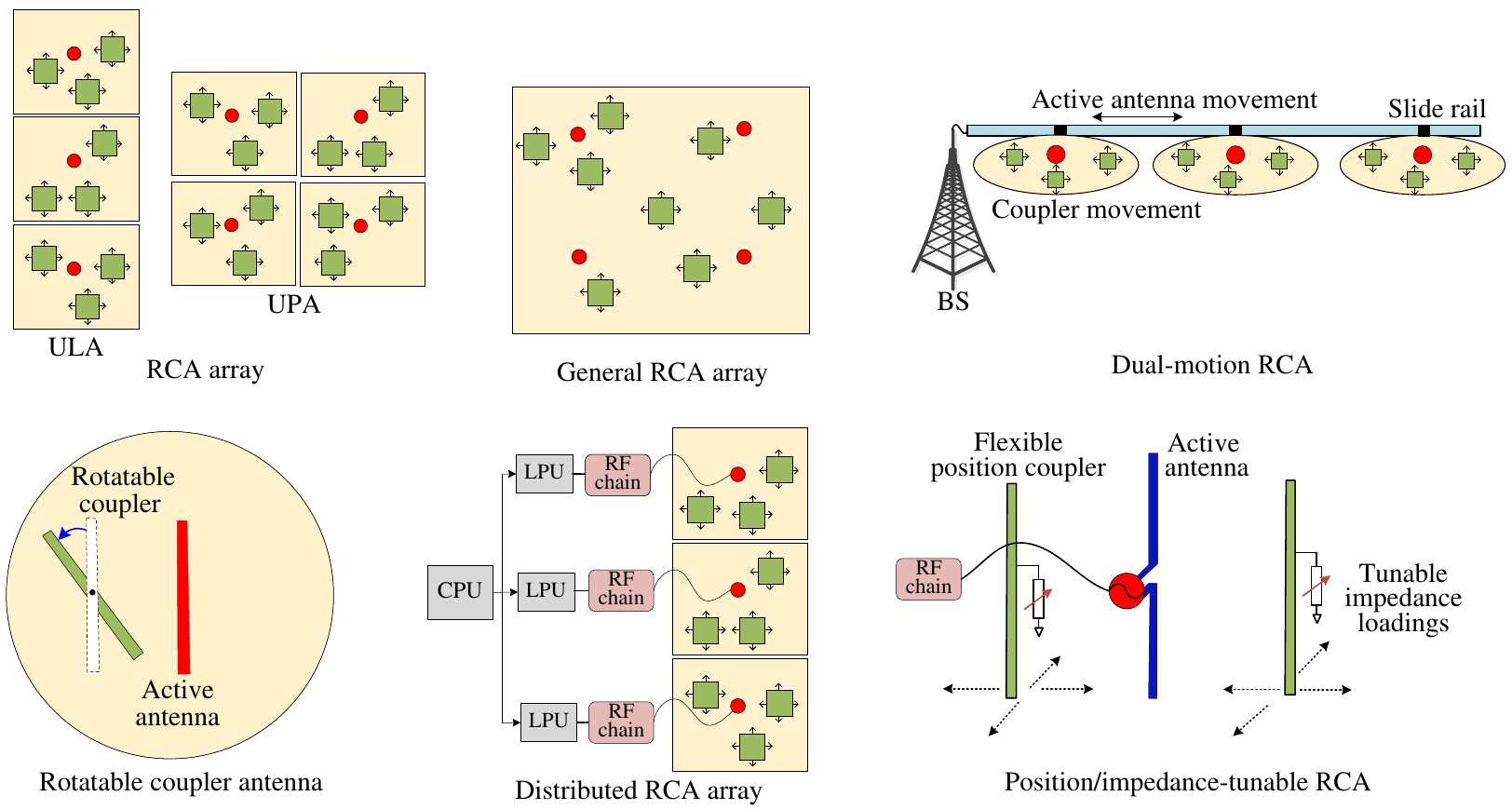}
	}
	\caption{RCA implementations in wireless networks.}
	\label{implement}
	\vspace{-0.59cm}
\end{figure*}

In addition to coupler translation, RCAs can reconfigure the induced currents by rotating the coupling elements.
As shown in Fig.~\ref{implement}, rotatable coupler antenna can be realized by compact rotary actuators, such as MEMS-based torsional actuators or miniature servo motors, mounted at the fixed centers of the couplers to enable independent 3D orientation adjustment \cite{ROCATWC}.
By rotating the passive couplers around the fixed active antenna, the radiation pattern can be reconfigured without moving the active antenna/RF chain \cite{ROCATWC}.

\subsection{Distributed RCA array}
In the RCA arrays as discussed, the received signals must be collected at the CPU to perform key communication tasks.
As the numbers of couplers and antennas grow, such centralized processing incurs high computational complexity and latency. To enhance performance while alleviating the computational burden at the CPU,
Fig.~\ref{implement} shows a distributed RCA array, which consists of multiple RCAs, each equipped with a local processing unit (LPU) that performs signal processing tasks such as channel estimation, precoding, and combining \cite{shao2026coupler}. The LPUs can operate in parallel and exchange information with the CPU for coordinated processing.

\subsection{Position/Impedance-Tunable RCA}
In the multiport circuit model, the coupler position and the load impedances jointly determine the induced current distribution on the couplers and, consequently, the mechanical beamforming weights. Therefore, a position- and impedance-tunable RCA is shown in Fig.~\ref{implement}, which jointly exploits coupler translation and electronic load-impedance $\mathbf{X}$ tuning \cite{DuRCAload}.
Each passive-coupler port is terminated by a tunable impedance load, which can be tuned using varactors.
Compared with position-only RCAs, the position-impedance RCA architecture enlarges the feasible induced-current excitation set. Consequently, the transmitter can achieve finer control of the induced-current magnitudes and phases for beam strengthening and interference suppression \cite{DuRCAload}.

\section{RCA Performance Enhancement}

\emph{Mechanical Beamforming Gain}: The RCAs can provide mechanical beamforming gain, which refers to the signal-to-noise ratio (SNR) or rate improvement achieved by mechanically reshaping the radiated field via passive-coupler reconfiguration.
In an RCA, the passive couplers radiate through induced currents generated by position-dependent EM coupling.
By tuning the phases and magnitudes of the induced excitations through mechanical beamforming to better match the channel responses, RCA can strengthen coherent combining, enable directional signal enhancement, and improve the effective channel gain, while complementing digital beamforming across active antennas in an RCA array.

\emph{Path-Loss Reduction}: The RCA technique can reduce large-scale path loss when the antenna translation range is non-negligible compared with the propagation distance.
For example, a dual-motion RCA can be mounted along the wall of a subway station, where it adapts the antenna position and radiation pattern to shorten the transmitter-receiver distance and overcome severe blockage caused by pillars, thereby maintaining reliable coverage.
This gain is particularly beneficial in blockage-prone environments, since it improves the dominant propagation condition rather than merely compensating for small-scale fading.

\emph{Fading Mitigation}: RCA reconfiguration mitigates deep fading by exploiting small-scale spatial channel variations within the local movement region. In multipath channels, slight changes in coupler position modify relative phases of dominant channel components, converting destructive combining into constructive combining. Therefore, RCAs avoid unfavorable channel realizations and reduce outage probability. This gain is challenging in high-mobility scenarios, since channel coherence time may be insufficient for coupler reconfiguration, and frequent coupler reconfiguration may incur non-negligible movement time/energy overhead. Hence, fading mitigation is most pronounced in slowly time-varying scenarios or when the coupler control is updated over a larger timescale based on statistical channel state information (CSI).

\emph{Spatial Multiplexing Gain}: The RCA technique can improve the spatial multiplexing gain of MIMO systems by employing RCAs at the transceiver.
Conventional MIMO systems may fail to achieve the full multiplexing gain due to rank deficiency caused by limited scattering and channel correlation.
In contrast, RCA apertures can be optimized to improve channel conditioning and yield more balanced singular values of the MIMO channel matrix, thereby increasing the achievable capacity.

\emph{Interference Suppression}: In multiuser scenarios, RCAs can enhance the desired-signal power while suppressing interference to and from undesired directions. For example, with RCA configurations designed at the BS based on long-term statistical CSI, subsequent digital precoding and combining based on instantaneous CSI become more effective for interference suppression.
This is because coupler repositioning can reduce inter-user channel correlation and facilitate spatial null formation with fewer RF chains.

\emph{Geometric Gain}: The RCAs can improve sensing and localization accuracy by exploiting geometric gain enabled by coupler position adjustability.
Target sensing accuracy depends on the relative geometry between the target and the transceiver apertures.
By adjusting coupler positions, the system can increase antenna geometric diversity, thereby yielding a geometric gain for detection and localization.
Determining RCA positions to exploit geometric gain under coupler movement constraints is an important direction for future work.

\section{Challenges and Potential Solutions}
\subsection{Coupler Position/Rotation Optimization}
One challenge to determine the coupler positions/rotations in RCAs lies in the nonconvex position constraints \cite{shao2026coupler,RCATWC,ROCATWC}. Moreover, the coupler positions/rotations are interdependent in both the channel response and the mechanical beamforming, which makes coupler-position optimization a complex task. A practical approach to tackle this challenge is to apply parallel block-coordinate updates combined with successive convex approximation to update coupler positions and rotations across different RCAs under local constraints, thus enabling distributed or parallel optimization with limited coordination overhead.

Although continuously tuning the coupler positions/rotations offers the highest flexibility and performance gains over fixed-position antennas, it is difficult to realize in practice because mechanical actuators typically support only discrete adjustment steps. For practical implementation, discrete position/rotation levels can be handled by continuous relaxation followed by quantization. This approach lowers the computational burden relative to exhaustive discrete search, but it introduces quantization errors. Hence, further research is required to design efficient approaches for obtaining discrete solutions.

\subsection{RCA Channel Estimation}
In RCA systems, channel estimation aims to obtain instantaneous or statistical CSI over the available coupler positions within a given transmitter or receiver region for enabling coupler position optimization. Unlike conventional MIMO channel estimation with a finite set of fixed antenna locations, RCA involves a much larger number of channels since couplers can move to arbitrary positions within the region.

Since couplers can only couple signals from the active antenna and cannot directly receive pilot signals, a low-pilot-overhead approach is to adopt a couplers-off scheme during channel training. In this case, only active antennas participate in pilot reception, and the active-antenna channels are estimated. Based on the active-antenna channels, the dominant path parameters can be extracted and used to reconstruct coupler channels for arbitrary coupler positions without additional pilots, where the position-dependent mutual impedance is obtained from a pre-established electromagnetic model. Hence, the pilot overhead mainly depends on the number of active antennas.
However, the couplers-off scheme may lead to severe fading of pilot signals when the active-antenna positions are poorly located, reducing channel reconstruction accuracy. To improve estimation accuracy, an approach is to keep the couplers on and move them over a number of random sample positions to collect data for reconstructing CSI for arbitrary coupler positions \cite{shao2026coupler}. Alternatively, active antenna mobility in the RCA system can be exploited to mitigate severe pilot fading.

\subsection{Coupler Trajectory Planning}
After obtaining the desired RCA configurations, a practical challenge is coupler trajectory scheduling, i.e., how to move each coupler from its initial location to a designated target location in a manner that is efficient in terms of both movement distance and movement delay.
Since reconfiguration inevitably induces transient channel variations, the trajectory planning problem should balance communication performance against movement time and energy cost.
Promising directions include hybrid parallel and sequential position updates across RCAs, coordination between transmitter and receiver reconfiguration, and a two-timescale operation in which coupler trajectories are optimized based on long-term statistical CSI, while fast-timescale resource allocation adapts to instantaneous CSI.

\section{Applications of Reconfigurable Coupler Antennas}
\begin{figure*}[t!]
	\centering
		\setlength{\abovecaptionskip}{0.cm}
	\fbox{\includegraphics[width=6.4in]{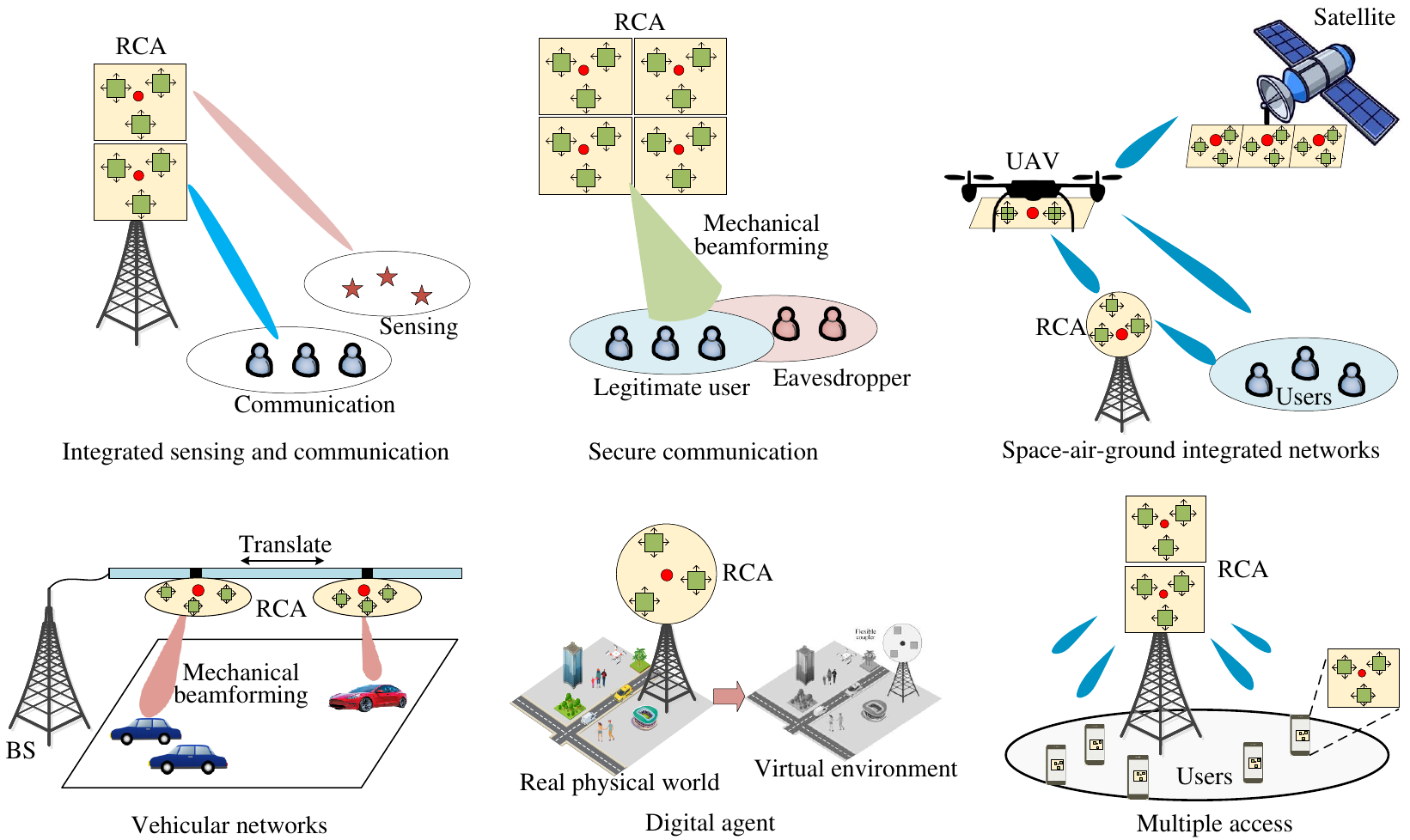}
	}
	\caption{Potential RCA applications in wireless networks.}
	\label{application}
	\vspace{-0.59cm}
\end{figure*}
We illustrate some potential applications of RCAs in wireless networks, which are depicted in Fig. \ref{application}.

\emph{Reconfigurable Coupler Antenna Aided ISAC}:
Integrated sensing and communication (ISAC) is expected to be a key component of 6G.
RCAs can establish previously unavailable LoS links or strengthen existing LoS paths, which is especially useful for ISAC deployments. In addition, RCA arrays can form distributed anchors with few RF chains, where enlarged inter-anchor spacing by moving RCAs improves geometric diversity for localization.

\emph{Reconfigurable Coupler Antenna Aided Secure Communication}:
Due to the broadcast nature of wireless transmission, confidential data is vulnerable to eavesdropping and jamming in dense and dynamic 6G deployments. By reconfiguring coupler positions, RCAs introduce additional spatial DoF for physical-layer security, which can be jointly optimized with digital beamforming, power allocation, and artificial noise to strengthen the legitimate link while degrading the eavesdropper's links. Open issues include secrecy analysis under spatially correlated fading and coupler movement constraints, and robust design when eavesdroppers can reconfigure their coupler/antenna positions.

\emph{Reconfigurable Coupler Antenna Aided Space-Air-Ground Integrated Network}:
In space-air-ground integrated networks (SAGIN), long satellite link propagation distances cause severe rate loss. RCAs are appealing for size, weight, and power (SWAP)-constrained satellite terminals to maintain link margin and mitigate interference. Moreover, RCAs can be mounted on unmanned aerial vehicles (UAVs) to enhance low-altitude communications \cite{FCjstsp}. Trajectory optimization for traditional UAVs incurs extra energy consumption and delay. In contrast, equipping UAVs with RCAs facilitates LoS establishment and mechanical beamforming with reduced UAV movement.

\emph{Reconfigurable Coupler Antenna Aided Vehicular Networks}:
The main challenges in vehicular networks arise from obstacles that can block LoS communication paths \cite{9186820}. To address these challenges, the active antennas of RCAs mounted on slide rails along roads and bridges can adjust their radiation locations toward vehicles to create or reinforce LoS links. In addition, coupler position control can broaden angular coverage and reduce infrastructure density.

\emph{Digital Agent and AI for Reconfigurable Coupler Antennas}:
The digital agent is appealing for RCAs because reconstructing position-dependent EM coupling in dynamic environments is expensive. By exploiting an EM map, it can predict channel statistics and generate long-term control policies, reducing CSI overhead and online computation. Moreover, RCAs can support future artificial intelligence (AI) services by improving connectivity for updates and synchronization, enabling high-throughput links, and reducing distributed edge intelligence latency.

\emph{Reconfigurable Coupler Antenna Aided Multiple Access}:
Conventional multiuser access designs are limited by fixed array geometries and RF chain costs, especially when closely spaced users exhibit highly correlated channels. Since couplers require no RF chains and have a small footprint, RCAs are suitable for both BSs and space-constrained terminals. In dense deployments, subwavelength coupler displacements at users or BSs can efficiently reshape the effective channels, improve channel conditioning, and suppress inter-user interference.

\section{Numerical Results}
\begin{figure*}[t!]
	\centering
		\setlength{\abovecaptionskip}{0.cm}
	\fbox{\includegraphics[width=6.99in]{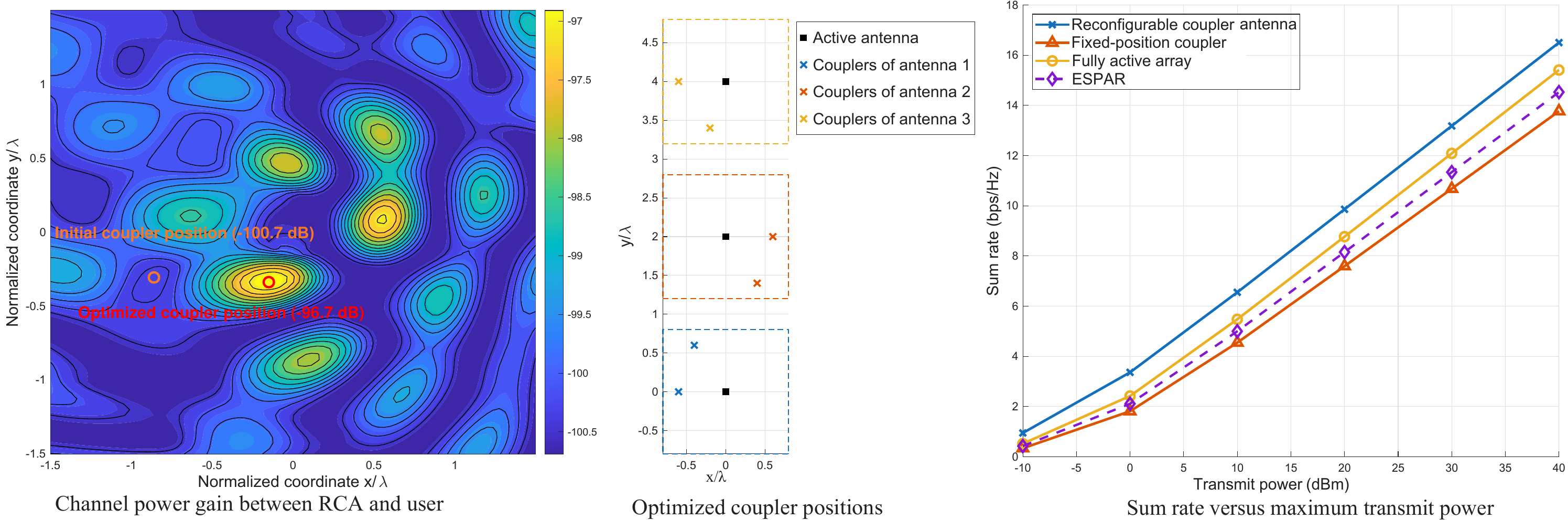}
	}
	\caption{RCA-aided MISO communication.}
	\label{simu}
	\vspace{-0.59cm}
\end{figure*}
We consider a downlink multiple-input single-output (MISO) communication system with a 13-path multipath channel serving one user. The BS is equipped with a ULA RCA array consisting of $M=3$ RCAs, each with $N=2$ couplers. The signal wavelength is set to $\lambda=0.04$ m. Each active antenna and each passive coupler are modeled as thin straight-wire dipoles of length $0.5 \lambda$.
We aim to maximize the received SNR of the RCA array-aided wireless system by optimizing the coupler positions using a parallel block-coordinate successive convex approximation algorithm \cite{shao2026coupler}. We compare the  RCA system with the following three benchmark schemes. 1) \emph{Fully active array:} All $MN+M$ ports are configured as active antennas at fixed positions with half-wavelength spacing.
2) \emph{Fixed-position couplers:} The couplers are fixed at uniformly spaced positions with an inter-element spacing of $0.4\lambda$.
3) \emph{ESPAR:} With the same fixed coupler positions as in fixed-position couplers, the load-impedance matrix is optimized using the method in \cite{10993454} to maximize the sum rate.

Fig.~\ref{simu} shows a realization of the channel power gain in dB as a function of a coupler position within its feasible movement region. The gain's irregular spatial distribution with multiple local extrema indicates that subwavelength coupler movement within a small feasible region can substantially enhance the effective channel. Furthermore, Fig.~\ref{simu} presents the optimized coupler locations around each fixed active antenna. Since the incoming waves arrive from multiple directions with unequal gains, the optimal mechanical beamforming weights and optimized coupler positions vary across RCAs to match the local channel distribution.
Moreover, Fig.~\ref{simu} shows the achievable rate versus the maximum transmit power for different schemes.
Under the considered MISO setup, the RCA array can outperform the fixed-geometry fully active array benchmark due to coupler position optimization and local spatial diversity gains.
The gain comes from coupler position optimization, which exploits both mechanical beamforming and spatial diversity. In addition, coupler repositioning can steer the signal link away from deep fading and enable a constructive combination of dominant paths.

\section{Conclusions}
In this article, we have discussed the RCA technique for enhancing wireless network performance by harnessing mutual coupling through coupler position/rotation reconfiguration.
We have introduced the basic channel characteristics, hardware architecture, and representative RCA implementations, and discussed key design issues for RCA-aided communications.
Notably, RCA-enabled transceivers can sense the wireless environment and dynamically adjust coupler positions to support diverse applications and services, by leveraging advanced optimization and signal processing algorithms.
We foresee that integrating RCAs into next-generation wireless systems can enhance antenna adaptability with a compact structure and low hardware overhead, and can open up new opportunities for applications and research.

\bibliographystyle{IEEEtran}
\bibliography{fabs}
\section*{Biographies}
\vspace{-35pt}
\begin{IEEEbiographynophoto}{Xiaodan Shao}
(x6shao@uwaterloo.ca) is a Postdoctoral Fellow with the Dept. of Electrical and Computer Engineering, University of Waterloo, Canada.
\end{IEEEbiographynophoto}
\vspace{-35pt}
\begin{IEEEbiographynophoto}{Weihua Zhuang}
(wzhuang@uwaterloo.ca) is a Professor with the Department of Electrical and Computer Engineering, University of Waterloo, Canada.
\end{IEEEbiographynophoto}
\vspace{-35pt}
\begin{IEEEbiographynophoto}{Xuemin Shen}
(sshen@uwaterloo.ca) is a Professor with the Department of Electrical and Computer Engineering, University of Waterloo, Canada.
\end{IEEEbiographynophoto}
\end{document}